\documentclass[preprint]{elsarticle}

\usepackage{lineno,hyperref,amsmath,graphicx}

\journal{Journal of Magnetism and Magnetic Materials}

\def\beq{\begin{equation}}
\def\eeq{\end{equation}}

\begin{document}

\begin{frontmatter}

\title{Model of charge triplets for high-T$_c$ cuprates}

\author[mainaddress,mysecondaryaddress]{A.~S. Moskvin\corref{cor}}
\ead{alexander.moskvin@urfu.ru}
\author[mainaddress]{Yu.~D. Panov}
\address[mainaddress]{Ural Federal University, 620083, Ekaterinburg, Russia}
\address[mysecondaryaddress]{Institute of Metal Physics UB RAS, 620108, Ekaterinburg, Russia}
\cortext[cor]{Corresponding author}

\begin{abstract}
Starting with a minimal model for the CuO$_2$ planes  with the on-site Hilbert space  reduced to a charge triplet of the three effective valence centers [CuO$_4$]$^{7-,6-,5-}$ (nominally Cu$^{1+,2+,3+}$) with different conventional spin,  different orbital symmetry, and different local lattice configuration, we develop a unified non-BCS spin-pseudospin model to describe the main phase states of doped cuprates.
We argue that antiferromagnetic insulating, charge ordered, superconducting, and Fermi-liquid phases are possible phase states of a model parent cuprate, while typical phase state of a doped cuprate, in particular mysterious pseudogap phase, is a result of a phase separation.
Superconductivity of cuprates is not a consequence of pairing of doped holes, but the result of  quantum transport of on-site composite hole bosons, whereas main peculiarities of normal state can be related to an electron-hole interplay for unusual Fermi-liquid phase and features of the phase separation.
Puzzlingly, but it is the electron-lattice interaction, which in the BCS model determines $s$-wave pairing, in the model of local composite bosons gives $d_{x^2-y^2}$-symmetry of the superconducting order parameter, thus showing once again a substantial involvement of the
lattice in the  cuprate's HTSC.

\end{abstract}

\begin{keyword}
HTSC cuprates, charge triplets \sep pseudospin, composite boson, phase separation
\end{keyword}

\end{frontmatter}


\section{Introduction}
Despite more than three decades that have passed since the discovery of HTSC\,\cite{HTSC}, today there is no consensus on the theoretical model that allows, within the framework of a single scenario, to describe the phase diagram of the high-$T_c$ cuprates, including  HTSC mechanism itself,  pseudogap phase, strange metal phase, a variety of static and dynamic fluctuations, etc.
 In our point
of view we miss several fundamental points: strong but specific electron-lattice effects, inapplicability of the Bardeen-Cooper-Schrieffer (BCS) paradigm which implies
 a search for a "superconducting glue"\, for the  $\bf k$-momentum pairing of metallic quasiparticles,
and inherent intrinsic electronic phase separation in cuprates.

 Recent precision measurements of various physical characteristics on thousands of cuprate samples\,\cite{Bozovic} indicate "insurmountable"\, qualitative and a few orders of magnitude quantitative discrepancies with ideas based on the canonical BCS approach and calculations based on Hubbard and t-J models, and rather support local real-space pairing mechanism for HTSC cuprates.
Numerous physicists argued that pairing is in fact local in cuprates, however, proposed mechanisms vary in many and sometime quite important details.
Appealingly simple and attractive picture of the preformed pairs and BEC superconductivity in cuprates, however, came to be at odds with a number of experimental observations of the typical Fermi liquid behavior, and notably with indications that
a well-defined Fermi surface (FS) exists, at least in overdoped cuprates, the thermal and electrical conductivity
were found to follow the standard Wiedemann-Franz law, quantum oscillations have been observed as well in various cuprates\,\cite{Bozovic}.
However, this contradictory behavior can be easily explained if we take into account the possibility of a phase separation, in particular, separating the superconducting BEC phase and the normal Fermi-liquid phase. Indeed, recently Pelc {\it et al}.\,\cite{Pelc} have introduced a phenomenological model wherein two electronic subsystems coexist within the unit cell: itinerant and localized holes, with the
$p$ holes introduced via doping always being itinerant while pairing is associated with the localized holes. Their minimalistic phenomenological model based on the localization/itineracy interplay and intrinsic electronic inhomogeneity captures key unconventional experimental results for the normal and superconducting state behavior at a quantitative level. The success and simplicity of the model greatly demystify the cuprate phase diagram and unambiguously point to a local superconducting pairing mechanism rather than to BCS. In fact, they argue that the Fermi liquid subsystem in cuprates is responsible for the normal state with angle-resolved photoemission spectra (ARPES), magnetic quantum oscillations, and Fermi arcs, but not for the unconventional superconducting state. In other words, {\it cuprate superconductivity is not related to the doped hole pairing}, the carriers which exhibit the Fermi liquid behaviour are not the ones that give rise to superconductivity.

Despite the fact that a large variety of different theoretical models has been designed to account for exotic electronic properties of cuprates and to shed light on their interplay with unconventional superconductivity, the most important questions remain unanswered to date.

In the paper we address a minimal model for the CuO$_4$-centers in CuO$_2$ planes  with the on-site Hilbert space  reduced to only three effective valence centers [CuO$_4$]$^{7-,6-,5-}$, or cluster analogs of Cu$^{1+,2+,3+}$ centers, forming a "well isolated"\, charge triplet\,\cite{truegap,dispro,Moskvin-JSNM-2019}. The very possibility of considering these centers on equal footing is predetermined by the strong effects of electron-lattice relaxation in cuprates\,\cite{Mallett,Moskvin-PSS-2020}.
Following the spin-magnetic analogy proposed by Rice and Sneddon\,\cite{Rice_1981},  we develop an  "unparticle"\, S\,=\,1 pseudospin model for the CuO$_2$ planes introducing an  effective spin-pseudospin Hamiltonian which takes into account both local and nonlocal correlations, single and two-particle transport,  Heisenberg spin exchange interaction, as well as electron-lattice coupling. As particular phase states we consider  antiferromagnetic insulator (AFMI), charge order  (CO), glueless $d$-wave Bose superfluid phase (BS), and unusual metallic phase (FL).

In full accordance with Hirsch's concepts of hole superconductivity\,\cite{Hirsch}
 we argue that the cuprate superconductivity is related with a composite on-site hole boson quantum transport, not with pairing of doped holes, in full accordance with the authors of Ref.\,\cite{Pelc}.

Puzzlingly, but it is the electron-lattice interaction, which in the BCS model determines $s$-wave pairing, in the model of local composite bosons gives $d_{x^2-y^2}$-symmetry of the superconducting order parameter, thus showing once again a substantial involvement of the lattice in the  cuprate's HTSC.

\section{Working model of the C$\mbox{u}$O$_4$ centers}
Despite the different crystal structure of undoped and electron/hole doped quasi-two-dimensional cuprates, their anomalous properties in the normal and superconducting states are associated with a common structural element, namely, the CuO$_2$ planes formed by the corner-shared CuO$_4$ clusters. As a first approximation the out-of-plane stuff can be considered as a source of the "external"\, crystal field,  which can have a strong effect on the values of the parameters of the effective Hamiltonian for the in-plane CuO$_4$-centers, in particular, on the ground state of the CuO$_2$ planes. The story of the superconductivity of the "parent" cuprate with the $T^{\prime}$-structure\,\cite{Naito-2016} makes us reconsider the very concept of the "parent cuprate", which is still associated with the undoped stoichiometric composition with the ground antiferromagnetic insulating state typical for cuprates with the $T$-structure. Below, by the "parent"\, we mean cuprate with the hole half-filling of the in-plane CuO$_4$ centers,  which, depending on the "external"\, crystal field, can have a different ground state, including an antiferromagnetic insulator, a superconductor, a Fermi metal,  nonmagnetic charge ordered insulator, or more intricate quantum state. Thus, we emphasize that the unusual properties of cuprates are largely the result of the "competition"\, of parameters that control the ground state of the CuO$_2$-planes.

The exclusion of the BCS mechanism as the main candidate for explaining the HTSC in cuprates does not mean excluding the important, if not decisive, role of the electron-lattice effects for explaining the unusual behavior of cuprates,  however, beyond the BCS theory.
In our opinion, the main effect of the electron-lattice interaction in cuprates is not the effect of Cooper pairing, but the effect of suppression of local and nonlocal electron correlations.
Thus, low-energy single-particle transfer  Cu$^{2+}$-\,Cu$^{2+}$$\rightarrow$\,Cu$^{1+}$-\,Cu$^{3+}$, or rather, [CuO$_4$]$^{6-}$+[CuO$_4$]$^{6-}$$\rightarrow$[CuO$_4$]$^{7-}$+[CuO$_4$]$^{5-}$ (11$\rightarrow$\,02, 20), under photoexcitation in parent cuprates, or Franck-Condon (FC) CT-transition, does create a bounded electron-hole (EH) pair, or CT exciton, and defines an optical CT gap $U_{opt}$, or effective local correlation parameter $U$, with a magnitude near 2.0\,eV (1.5\,eV), typical for all the undoped cuprates with the $T$-type ($T^{\prime}$) structure characterized by the presence (absence) of apical oxygen above and below the CuO$_2$ planes.
However, after photoexcitation, the strong electron-lattice polarization effects lead to a relaxation of these CT excitons  with the formation of localized low-energy metastable polaronic-like EH dimers, or, more precisely, coupled electron ([CuO$_4^*$]$^{7-}$) and hole ([CuO$_4^*$]$^{5-}$) centers (02, 20), where the star points to a relaxed local lattice surroundings. Charge (electron/hole) transfer in cuprates is accompanied by strong local lattice deformations: for CuO$_4$-centers in cuprates the Cu-O separation increases by 0.1\,\AA\, from Cu$^{3+}$ to Cu$^{2+}$ and another 0.1\,\AA\, from Cu$^{2+}$ to Cu$^{1+}$\,\cite{Larsson_2007}.
The electron-lattice interaction
 leads to the stability of the electron and hole centers in the lattice of the parent cuprate, however, the ground states of all three centers -- the electron, parent, and hole one  will correspond to different values of the local breathing configuration coordinate $Q_{A_{1g}}$: $+Q_0$, 0, $-Q_0$, respectively.

The puzzling near-degeneracy of bare (11) and relaxed CT-states 20 (02), or the disproportionation instability in cuprates which has been demonstrated in accurate quantum-chemical calculations\,\cite{Larsson_2007} is a peculiar feature of all the parent cuprates.
  Different experimental data point to $U_{th}$\,$\approx$\,0.4\,eV for $T$-La$_2$CuO$_4$\,\cite{Gorkov,Gorkov1,truegap}, while for parent $T^{\prime}$-cuprates this value can be close to zero, or even negative\,\cite{CM_2021}.

Small magnitude of the EH-dimer formation energy $U_{th}$, or a "thermal"\, (adiabatic) CT gap results in instability regarding the charge transfer
with the disproportionation and formation of the system both of coupled and individual electron and hole centers. In other words, all the three charge centers
 [CuO$_4^*$]$^{7-}$, [CuO$_4$]$^{6-}$, [CuO$_4^*$]$^{5-}$, or cluster analogs of Cu$^{1+,2+,3+}$ centers,  must be considered on equal footing.

Our scenario for 2D cuprates is based on a minimal model for the CuO$_4$-centers in CuO$_2$ planes  with the on-site Hilbert space  reduced to only these three effective valence centers, forming a "well isolated"\, charge triplet\,\cite{truegap,dispro,Moskvin-JSNM-2019}.
 These centers to be many-electron atomic species with strong $p-d$ covalence and strong intra-center correlations cannot be described within any conventional (quasi)particle approach. We combine the three centers into a pseudospin $S$\,=\,1 triplet (see Table\,1) following the spin-magnetic analogy proposed by Rice and Sneddon\,\cite{Rice_1981} to describe the three charge states (Bi$^{3+}$, Bi$^{4+}$, Bi$^{5+}$) of the bismuth ion in  BaBi$_{1-x}$Pb$_x$O$_3$ and use the traditional spin algebra.

The centers are characterized by different conventional spin: $s$\,=\,1/2 for "parent"\, [CuO$_4$]$^{6-}$ center and $s$\,=\,0 for electron and hole centers ([CuO$_4^*$]$^{7-}$- and [CuO$_4^*$]$^{5-}$-centers) and different orbital symmetry:$B_{1g}$ for the ground states  of the  [CuO$_4$]$^{6-}$ center,  $A_{1g}$   for the electron and hole centers, and different local configurational coordinate (see Table\,\ref{table}).

\begin{table*}
\begin{center}
\caption{Specific features of the charge triplet in cuprates, Q is configurational coordinate.}
\begin{tabular}{|c|c|c|c|c|c|c|}
\hline
 Center & Cluster & Nominal & Pseudospin S=1 projection & Spin & Orbital state & Q  \\ \hline
electron & [CuO$_4^*$]$^{7-}$ & Cu$^{1+}$ & M$_S$\,=\,{--}1 & 0 & $A_{1g}$ & +Q$_0$ \\ \hline
parent & [CuO$_4$]$^{6-}$ & Cu$^{2+}$ & M$_S$\,=\,0 & 1/2 & $B_{1g}$ & 0 \\ \hline
hole & [CuO$_4^*$]$^{5-}$ & Cu$^{3+}$ & M$_S$\,=\,{+}1 & 0 & $A_{1g}$ & -Q$_0$ \\ \hline
\end{tabular}
\label{table}
\end{center}
\end{table*}

\section{Charge triplet model: S\,=\,1 pseudospin formalism}
To describe the diagonal and off-diagonal, or quantum local charge order we start with a simplified {\it charge triplet model}
that implies a full neglect of spin and orbital degrees of freedom\,\cite{truegap,dispro,Moskvin-JSNM-2019}.
Three charge states of the CuO$_4$ plaquette: a bare center $M^0$=[CuO$_4$]$^{6-}$, a hole center $M^{+}$=[CuO$_4^*$]$^{5-}$,
and an electron center $M^{-}$=[CuO$_4^*$]$^{7-}$ are assigned to  three components of the $S$\,=\,1 pseudospin triplet
with the pseudospin projections  $M_S =0,+1,-1$, respectively.

Pseudospin formalism is one of the most promising "unparticle"\, approaches. This approach allows one to use many results that are well known for spin-magnetic systems, including a description of phase transitions, topological structures, domains, and domain walls\,\cite{Batista}.

The $S$\,=\,1 spin algebra includes the eight independent nontrivial pseudospin operators, the three dipole and five quadrupole operators\,\cite{Moskvin-SCES,Moskvin-JETP-2015,Moskvin-JSNM-2018,Moskvin-JSNM-2019} with irreducible form as follows
\begin{equation}
	{\hat S}_z;\, {\hat S}_{\pm}=\frac{1}{\sqrt{2}}({\hat S}_{x}\pm i{\hat S}_{y});\,{\hat S}_z^2;\,{\hat T}_{\pm}=\{{\hat S}_z, {\hat S}_{\pm}\};\,{\hat S}^2_{\pm} \, .
\end{equation}
with hermitian conjugation as follows
$$
{\hat S}_z^{\dagger}={\hat S}_z, ({\hat S}_z^2)^{\dagger}={\hat S}_z^2, {\hat S}_{\pm}^{\dagger}={\hat S}_{\mp}, {\hat T}_{\pm}^{\dagger}={\hat T}_{\mp}, ({\hat S}_{\pm}^2)^{\dagger}={\hat S}_{\mp}^2 \,.
$$

The two Fermi-like  pseudospin raising/lowering operators
${\hat S}_{\pm}$ and ${\hat T}_{\pm}$  change the pseudospin projection by $\pm 1$, with slightly different properties
\begin{eqnarray}
\langle 0 |\hat S_{\pm} | \mp 1 \rangle = \langle \pm 1 |\hat S_{\pm} | 0
\rangle =1 ;\,
\langle 0 |\hat T_{\pm}| \mp 1 \rangle = -\langle \pm 1 |(\hat T_{\pm}| 0 \rangle =\mp 1.
\end{eqnarray}
In lieu of ${\hat S}_{\pm}$ and ${\hat T}_{\pm}$ operators one may use  two novel operators:
$$
	{\hat P}_{\pm}=\frac{1}{2}({\hat S}_{\pm}+{\hat T}_{\pm});\,{\hat N}_{\pm}=\frac{1}{2}({\hat S}_{\pm}-{\hat T}_{\pm})\,,
$$
which do realize transformations [CuO$_4$]$^{6-}$$\leftrightarrow$\,[CuO$_4$]$^{5-}$ and [CuO$_4$]$^{6-}$$\leftrightarrow$\,[CuO$_4$]$^{7-}$, respectively.

Strictly speaking, we should extend the on-site Hilbert space to a spin-pseudospin quartet $|SM;s\nu\rangle$: $|1\pm 1;00\rangle$ and $|10;\frac{1}{2}\nu\rangle$, where $\nu =\pm\frac{1}{2}$, and instead of spinless operators ${\hat P}_{\pm}$ and ${\hat N}_{\pm}$ introduce operators ${\hat P}_{\pm}^{\nu}$ and ${\hat N}_{\pm}^{\nu}$, which transform both on-site charge (pseudospin) and spin states as follows
$$
\hat{P}_{+}^{\nu} | 10; \frac{1}{2} \, {-} \nu \rangle = | 11; 00 \rangle ; \,
\hat{P}_{-}^{\nu} | 11; 00 \rangle = | 10; \frac{1}{2} \, {-} \nu \rangle ;
$$
\begin{equation}
\hat{N}_{+}^{\nu} | 1\, {-}1; 00 \rangle =| 10; \frac{1}{2} \, \nu \rangle ; \;
\hat{N}_{-}^{\nu} | 10; \frac{1}{2} \, \nu \rangle = | 1\, {-}1; 00 \rangle .
\end{equation}
These Fermi-like operators obey the anticommutation permutation rules.

It should be noted that inclusion of conventional spin-1/2 moment operator $\hat{\bf s}=\frac{1}{2}\boldsymbol{\sigma}$  allows one to establish a one-to-one correspondence between all the spin-pseudospin operators and the on-site Hubbard $X$-operators\,\cite{Hubbard}: ${\hat X}^{\alpha\beta}=|\alpha\rangle \langle\beta |$ acting in the space of the four eigenstates  $|0\rangle =|1{-}1;00\rangle$,  $|\sigma\rangle =|10;\frac{1}{2}\sigma\rangle$, $|2\rangle =|11;00\rangle$
$$
{\hat S}_z={\hat X}^{22}-{\hat X}^{00},{\hat S}^2_z={\hat X}^{22}+{\hat X}^{00},{\hat S}^2_{+}={\hat X}^{20},{\hat S}^2_{-}={\hat X}^{02},
$$
\begin{equation}
1-{\hat S}^2_z={\hat X}^{\sigma\sigma},{\hat P}_{+}={\hat X}^{2\sigma},{\hat P}_{-}={\hat X}^{\sigma 2},
{\hat N}_{+}={\hat X}^{0\sigma},{\hat N}_{-}={\hat X}^{\sigma 0}.
\end{equation}
It means that algebra of the spin-pseudospin operators coincides with that of Hubbard ${\hat X}$-operators.

The two Bose-like  pseudospin raising/lowering operators
${\hat S}_{\pm}^{2}$  change the pseudospin projection by $\pm 2$, respectively. In other words, these operators can be associated with  creation/annihilation of an on-site hole pair, or composite on-site boson, with a kinematic constraint $({\hat S}_{\pm}^{2})^2$\,=\,0, that underlines its "hard-core"\, nature.
For the Bose creation/annihilation operators ${\hat B}^{\dag}$\,=\,${\hat S}_{+}^{2}$/${\hat B}$\,=\,${\hat S}_{-}^{2}$ we arrive at Fermi-like on-site and Bose-like inter-site permutation relations on the  space with the on-site $M_S=\pm 1$ states:
\beq
\{{\hat B}_i, {\hat B}^{\dag}_i\}=1\,\,, [{\hat B}_i, {\hat B}^{\dag}_j]=1 \, .
\eeq
The on-site anticommutation relation can be rewritten as follows
\beq
[{\hat B}_i, {\hat B}^{\dag}_i]=1-2{\hat B}^{\dag}_i{\hat B}_i=1-2{\hat N}_i\, .
\eeq
  The on-site local composite boson is composed of the strongly correlated spin singlet hole pair formally described by an orbital wave function
with $A_{1g}\propto d_{x^2-y^2}^2$ symmetry ("extended $s$\,-wave") and can be called the on-site Zhang-Rice boson.
Simplified, the local on-site composite hole boson is a pair of holes coupled by local correlations both with each other and with the "core", that is, the electronic center [CuO$_4^*$]$^{7-}$ (nominally Cu$^{1+}$). In fact, such a local boson exists only as an indivisible part of the ZR hole center [CuO$_4^*$]$^{5-}$ (nominally Cu$^{3+}$).
 In our point
of view, it is these on-site local two-hole bosons that are superconducting carriers both in hole- and electron-doped cuprates in full accordance with Hirsch's concepts of hole superconductivity\,\cite{Hirsch}. We emphasize the fundamental difference between such a local composite boson as a "preformed"\, pair and a Cooper pair composed of +${\bf k}$ and -${\bf k}$ partners. However, the on-site ZR boson also differs significantly from composite inter-site bosons, such as magnetic or lattice bipolarons\,\cite{Mueller,Alexandrov} or magnetic  "pairons"\,\cite{Sacks}.

The  boson-like  pseudospin raising/lowering operators
${\hat S}_{\pm}^{2}$  define a local pseudospin "nematic"\, order parameter
\begin{equation}
\langle {\hat S}_{\pm}^{2} \rangle\,=\,\frac{1}{2}(\langle {\hat S}_x^2-{\hat S}_y^2\rangle \pm i\langle\{{\hat S}_x,{\hat S}_y\}\rangle ) = |\langle {\hat
S}_{\pm}^{2} \rangle|e^{\pm 2i\alpha} \, .
\end{equation}
 Obviously, the  mean value $\langle {\hat S}_{\pm}^{2} \rangle$ can be addressed to be a complex superconducting local order parameter\,\cite{truegap}.

Instead of irreducible components, we can use the Cartesian components of the pseudospin operators\,\cite{Panov2019}
\begin{eqnarray}
	 {\hat S}_{\pm}^{2} \,=\,\frac{1}{2}\left(({\hat S}_x^2-{\hat S}_y^2) \pm i\{{\hat S}_x,{\hat S}_y\} \right)={\hat B}_1\pm i{\hat B}_2 \, ; \\ \nonumber
{\hat P}^{\nu}_{\pm}=\frac{1}{2}({\hat P}_1^{\nu}\pm i{\hat P}_2^{\nu}); \,{\hat N}^{\nu}_{\pm}=\frac{1}{2}({\hat N}_1^{\nu}\pm i{\hat N}_2^{\nu})
\end{eqnarray}
with hermitian operators ${\hat B}_{1,2}$, ${\hat P}_{1,2}^{\nu}$, ${\hat N}_{1,2}^{\nu}$

\section{Effective spin-pseudospin Hamiltonian}

As for conventional spin-magnetic systems, we can integrate out the high-energy degrees of freedom, and after projecting onto the Hilbert basis of well isolated charge triplet we have chosen, to arrive at the effective spin-pseudospin Hamiltonian obeying the spin and pseudospin kinematic rules.

Within a rigid lattice approximation the effective S\,=\,1 pseudospin Hamiltonian which does commute with the $z$-component of the total pseudospin  $\sum_{i}S_{iz}$ thus conserving the total charge of the system can be written to be a sum of potential and kinetic energies:
\begin{equation}
{\hat H}={\hat H}_{pot}+{\hat H}_{kin}\, .
\label{H}	
\end{equation}

Potential energy can be written as follows
\begin{equation}
	{\hat H}_{pot} =  \sum_{i}  (\Delta _{i}{\hat S}_{iz}^2
	  - \mu {\hat S}_{iz}) + \sum_{i>j} V_{ij}{\hat S}_{iz}{\hat S}_{jz}+ \sum_{i>j} v_{ij}{\hat S}^2_{iz}{\hat S}^2_{jz}\, ,
\label{Hch}	
\end{equation}
with a charge density constraint: $\frac{1}{N}\sum _{i} \langle {\hat S}_{iz}\rangle =n$,
where $n$ is the deviation from a half-filling.
The first on-site term in ${\hat H}_{pot}$, resembling single-ion spin anisotropy, describes the effects of a bare pseudospin splitting, or the local energy of $M^{0,\pm}$ centers and relates with the on-site density-density interactions, $\Delta$\,=\,$U$/2, $U$ being the local correlation parameter, or pair binding energy for composite boson. The second term   may be
related to a   pseudo-magnetic field $\parallel$\,$Z$ with $\mu$ being the hole chemical potential.  The third and fourth terms in ${\hat H}_{pot}$ describe the inter-site density-density interactions, or nonlocal correlations, with the third directly related to the Coulomb attraction/repulsion, while the fourth term can be called the covalent correlation correction.

Kinetic energy ${\hat H}_{kin}={\hat H}_{kin}^{(1)}+{\hat H}_{kin}^{(2)}$ is a sum of one-particle and two-particle transfer contributions.
In terms of Fermi-like spin-dependent operators ${\hat P}_{\pm}^{\nu}$ and  ${\hat N}_{\pm}^{\nu}$ the  Hamiltonian ${\hat H}_{kin}^{(1)}$ reads as follows:
\begin{equation}
{\hat H}_{kin}^{(1)}= -\sum_{i>j}\sum_{\nu} [t^p_{ij}{\hat P}_{i+}^{\nu}{\hat P}_{j-}^{\nu}+
 t^n_{ij}{\hat N}_{i+}^{\nu}{\hat N}_{j-}^{\nu}+
 \frac{1}{2} t^{pn}_{ij}({\hat P}_{i+}^{\nu}{\hat N}_{j-}^{\nu}+{\hat P}_{i-}^{\nu}{\hat N}_{j+}^{\nu}) +h.c.] \,,
\label{Hkin1}	
\end{equation}
where $t^p, t^n, t^{pn}$ are transfer integrals for the correlated hopping.
All the three terms here  suppose a clear physical interpretation. The first $PP$-type term describes one-particle transfer processes:
 Cu$^{3+}$+Cu$^{2+}$$\leftrightarrow$ Cu$^{2+}$+Cu$^{3+}$,
that is  a rather conventional  motion of the hole $M^+$ centers in the lattice formed by parent $M^0$ (Cu$^{2+}$)centers ($p$-type carriers, respectively) or the motion of the $M^0$ centers in the lattice formed by hole $M^+$ centers ($n$-type  carriers, respectively).
The second $NN$-type term describes
one-particle transfer processes: Cu$^{1+}$+Cu$^{2+}$$\leftrightarrow$ Cu$^{2+}$+Cu$^{1+}$,
that is  a rather conventional  motion of the electron $M^-$ centers in the lattice formed by $M^0$ centers ($n$-type carriers)
or the motion of the $M^0$ centers in the lattice formed by electron $M^-$ centers ($p$-type  carriers).
The third $PN$ ($NP$) term in ${\hat H}_{kin}^{(1)}$ defines a very different one-particle transfer process:
Cu$^{2+}$+Cu$^{2+}$$\leftrightarrow$ Cu$^{3+}$+Cu$^{1+}$, Cu$^{1+}$+Cu$^{3+}$,
that is the \emph{local disproportionation/recombination}, or the \emph{electron-hole pair creation/annihilation}. It is this interaction that is believed to be responsible for the violation of the "classical"\, Fermi-particle behavior.
Interestingly, the term can be related with a local pairing as  the hole $M^+$-center can be addressed to be a hole pair (=\,composite hole boson) localized on the  electron $M^-$-center.

Hamiltonian ${\hat H}_{kin}^{(2)}$:
\begin{equation}
  {\hat H}_{kin}^{(2)}=-\sum_{i>j} t_{ij}^b({\hat S}_{i+}^{2}{\hat S}_{j-}^{2}+{\hat S}_{i-}^{2}{\hat S}_{j+}^{2})\,,
  \label{H2}
\end{equation}
describes the two-particle (local composite boson) inter-site
  transfer, that is the  motion of the hole center in the lattice formed by the electron centers, or the exchange reaction:
Cu$^{3+}$+Cu$^{1+}$ $\leftrightarrow$ Cu$^{1+}$+Cu$^{3+}$.
In other words, $t^b_{ij}$ is the transfer integral for the on-site ZR boson.
Depending on the sign of $t^b$, this interaction will stabilize the superconducting $\eta_0$- ($t^b>0$)  or $\eta_{\pi}$- ($t^b<0$) phase.

Conventional Heisenberg spin exchange Cu$^{2+}$--Cu$^{2+}$ coupling should be transformed  as follows
\begin{equation}
{\hat H}_{ex}=\sum_{i>j} J_{ij}(\hat {\bf s}_i\cdot \hat {\bf s}_j)	\Rightarrow	{\hat H}_{ex}=s^2\sum_{i>j}J_{ij} (\boldsymbol{\sigma}_i\cdot  \boldsymbol{\sigma}_j)\, ,
\end{equation}
where operator $\boldsymbol{\sigma}=2\hat{\rho}^s \mathbf{s}$  takes into account the on-site spin density $\hat{\rho}^s=(1-{\hat S}_{z}^2)$.

The inclusion of spin exchange in the effective spin-pseudospin Hamiltonian requires additional comment.
Indeed, disregarding the effects of electron-lattice relaxation, it is the $PN$-type charge transfer [CuO$_4$]$^{6-}$+[CuO$_4$]$^{6-}$$\rightarrow$ [CuO$_4$]$^{5-}$+[CuO$_4$]$^{7-}$ through the 2\,eV optical charge transfer gap that contributes to the spin exchange with an exchange integral on the order of 0.1\,eV. However, the "unrelaxed" centers [CuO$_4$]$^{5-}$ and [CuO$_4$]$^{7-}$  have the high-energy local lattice configuration of the parent center [CuO$_4$]$^{6-}$. After taking into account the electron-lattice relaxation, we arrive at new "relaxed" centers [CuO$_4^*$]$^{5-}$ and [CuO$_4^*$]$^{7-}$ with a new local configuration and much lower energy, the optical gap is reduced to a thermal (adiabatic) gap with a much smaller value. The relaxed charge states are included in the "well-isolated" triplet, so that the spin exchange, as a result of taking into account the contribution of high-lying states, is included in the effective Hamiltonian.

Obviously, the spin exchange provides an energy gain to the parent antiferromagnetic insulating (AFMI) phase with $\langle {\hat S}_{iz}^2\rangle$\,=\,0, while local superconducting order parameter $\langle {\hat S}_{i\pm}^2\rangle$  is maximal given $\langle {\hat S}_{iz}^2\rangle$\,=\,1. In other words, the superconductivity and magnetism are nonsymbiotic phenomena with competing local order parameters giving rise to an intertwinning, glassiness, and other forms of electronic heterogeneities. In contrast to the main ideas of the spin-fluctuation HTSC scenario the conventional spin degree of freedom seems to play merely negative effect, magnetism is incompatible with optimal high-T$_c$ superconductivity.

Pseudospin formalism for CuO$_4$-centers allows one to give an effective description of the local lattice energy and local electron-lattice interaction. Indeed, in the general case, the electron-lattice interaction has a complex form, however,  in our point
of view, the most important role is played by the on-site (intra-center) interaction with the participation of a breathing mode  $Q(A_{1g})$ and two in-plane rhombic modes of the oxygen displacements,  $Q(B_{1g})$ and $Q(B_{2g})$ with symmetry $B_{1g}\propto d_{x^2-y^2}$ and $B_{2g}\propto d_{xy}$, respectively. The relevant pseudospin-lattice Hamiltonian  conserving the total charge of the system can be written as follows
$$
{\hat H}_{e-l}={\hat H}_{e-l}^{(A)}+{\hat H}_{e-l}^{(B)} \, ,
$$
where
\begin{equation}
{\hat H}_{e-l}^{(A)}=\sum_i(a_1{\hat S}_{iz}+a_2{\hat S}_{iz}^{2})Q_i(A_{1g}) \, ,
\end{equation}
$$
{\hat H}_{e-l}^{(B)} =b_1\sum_i({\hat S}_{i+}^{2}+{\hat S}_{i-}^{2}){\bf Q}_i(B_{1g})-ib_2\sum_i({\hat S}_{i+}^{2}-{\hat S}_{i-}^{2}){\bf Q}_i(B_{2g})=
$$
$$
b_1\sum_i({\hat S}_{ix}^{2}-{\hat S}_{iy}^{2}){\bf Q}_i(B_{1g})+b_2\sum_i\{\hat S_{ix},\hat S_{iy}\}{\bf Q}_i(B_{2g})=
$$
\begin{equation}
b_1\sum_i{\hat B_{1i}}{\bf Q}_i(B_{1g})+b_2\sum_i{\hat B_{2i}}{\bf Q}_i(B_{2g}) \, ,
\end{equation}
where $a_{1,2}, b_{1,2}$ are parameters of the intra-center electron-lattice interaction. Here we assume classical modes $Q(A_{1g})$, ${\bf Q}(B_{1g})$ and ${\bf Q}(B_{2g})$, where the vector form of the $B_{1g,2g}$-modes reflects two possible directions of polarization of these modes. It is worth noting that the breathing $A_{1g}$-mode includes the apex oxygen displacements. Obviously, ${\hat H}_{e-l}^{(2)}$ defines the in-plane ($xy$-) pseudospin anisotropy and plays a fundamental role in determining the symmetry of the superconducting order parameter in cuprates and specific behavior of bond-stretching phonon modes at superconducting transition.

Obviously, a complete description of the system should include the local elastic energy  which can be written in a simplified form as follows
$$
{\hat H}_{lat}=\frac{1}{2}\sum_{i,j}(K_{ij}(A_{1g})Q_i(A_{1g})Q_j(A_{1g})+
$$
\begin{equation}
K_{ij}(B_{1g}){\bf Q}_i(B_{1g}){\bf Q}_i(B_{1g})+K_{ij}(B_{2g}){\bf Q}_i(B_{2g}){\bf Q}_i(B_{2g}))\,,
\end{equation}
where we limited ourselves to the contribution of the actual modes of local displacements, $K_{ij}$ are elastic constants.
 One must take into account the presence of the two symmetry equivalent but orthogonal orientations of rhombic modes ${\bf Q}(B_{1g})$ and ${\bf Q}(B_{2g})$ and the possibility of their staggered ordering\,\cite{Mueller}, so that the elastic energy is minimized by the complimentary distortions.

 Note that the $A_{1g}$- and $B_{1g}$-modes are the  strongest phonon modes for cuprates, namely these are strongly correlated with the onset of superconductivity\,\cite{Bianconi,Oyanagi}.


\subsection{"Cartesian"\, form of the spin-pseudospin Hamiltonian}

Using the "Cartesian"\, form of the spin-pseudospin operators  one can
rewrite the spin-pseudospin Hamiltonian ${\hat H}$ in  an equivalent symbolic "vector"\, form as follows
$$
	\mathcal{H} = \Delta \sum_i {\hat S}_{zi}^2
	+ V \sum_{\left\langle ij\right\rangle} {\hat S}_{zi} {\hat S}_{zj}
	+ Js^2 \sum_{\langle ij \rangle} \boldsymbol{\hat \sigma}_i \boldsymbol{\hat \sigma}_j -
$$
$$
 \mu \sum_i {\hat S}_{zi} 	
	- \frac{t_b}{2} \sum_{\langle ij \rangle} \mathbf{{\hat B}}_{i} \mathbf{{\hat B}}_{j}
	- \frac{t_p}{2} \sum_{\langle ij \rangle \nu} \mathbf{\hat P}_{i}^{\nu} \mathbf{\hat P}_{j}^{\nu} -
$$
\begin{equation}
	 \frac{t_n}{2} \sum_{\langle ij \rangle \nu} \mathbf{\hat N}_{i}^{\nu} \mathbf{\hat N}_{j}^{\nu}
	- \frac{t_{pn}}{4} \sum_{\langle ij \rangle \nu}
		\left( \mathbf{\hat P}_{i}^{\nu} \mathbf{\hat N}_{j}^{\nu} + \mathbf{\hat N}_{i}^{\nu} \mathbf{\hat P}_{j}^{\nu} \right) \, ,
\label{HH}
\end{equation}
where we limited ourselves to the interaction of the nearest neighbors, $\boldsymbol{\hat \sigma}= ({\hat \sigma}_x, {\hat \sigma}_y, {\hat \sigma}_z)$, $\mathbf{{\hat B}}=({\hat B}_1, {\hat B}_2)$, $\mathbf{\hat P}^{\nu}=({\hat P}^{\nu}_1, {\hat P}^{\nu}_2)$, $\mathbf{\hat N}^{\nu}=({\hat N}^{\nu}_1, {\hat N}^{\nu}_2)$.

\section{Large variety of phases described by spin-pseudospin Hamiltonian}

Similar to conventional spin systems the $S$\,=\,1 pseudospin formalism allows us to predict various types of  diagonal and off-diagonal long-range order and pseudospin excitations, including commensurate and incommensurate charge orders (pseudospin density waves), superfluid and supersolid phases, different topological excitations typical for 2D systems\,\cite{Moskvin-JETP-2015,Moskvin-JSNM-2018,Moskvin-JSNM-2019}. The wide variety of possible phases is associated with a large number of local spin and pseudospin order parameters.

Neglecting the conventional spin degree of freedom the $S$\,=\,1 pseudospin Hamiltonian (\ref{H}) describes an extended bosonic Hubbard model (EBHM) with truncation
 of the on-site Hilbert space to the three lowest occupation states n = 0,\,1,\,2, or the model of semi-hard-core bosons\,\cite{Moskvin-JETP-2015}.
 The EBHM Hamiltonian is a paradigmatic model for the highly topical field of ultracold gases in optical lattices and also this is one of the working models to describe the insulator-metal transition and high-temperature superconductivity.

Interestingly, the pseudospin Hamiltonian  (\ref{H}), with the exception of the $ST$-term in Hamiltonian, which is not invariant with respect to time reversal, can be considered as a fairly general anisotropic non-Heisenberg Hamiltonian of the spin-magnetic $S$\,=\,1 system.
Currently, these systems are actively studied using various methods, from molecular field theory to the Green's function method for the Hubbard operators (see, e.g.,\,Ref.\cite{Fridman}).

Among the simple phases of the effective spin-pseudospin  Hamiltonian   with a single nonzero local order parameter ("single-OP"\, phases, or "monophases"),
we note the antiferromagnetic insulating (AFMI) phase with $\langle \boldsymbol{\hat \sigma}_i\rangle \not=0$, the charge ordering (CO) phase with $\langle {\hat S}_{zi}\rangle \not=0$,
and the bosonic superconducting (BS) phase with  $\langle {\hat S}_{\pm i}^2\rangle \not=0$.
It should be emphasized that the superconducting phase of cuprates is not a consequence of pairing of doped holes, but represents one of the possible phase states of parent cuprates.
All the   phases with long-range order we address above, AFMI, CO, BS are characterized by a nonzero local order parameters and can be called N\'eel phases which are typical for spin-magnetic systems.

Among phases with two nonzero local order parameters, we note the supersolid phase with $\langle {\hat S}_{zi}\rangle \not=0$ and $\langle {\hat S}_{\pm i}^2\rangle \not=0$, the phase of the spin-pseudospin wave with $\langle \boldsymbol{\hat \sigma}_i\rangle \not=0$ and  $\langle {\hat S}_{zi}\rangle \not=0$.
Taking into account the electron-lattice interaction will inevitably lead to the fact that both charge ordering and superconductivity will be accompanied by structural distortions.
Furthermore, the inclusion of the electron-lattice interaction ${\hat H}_{e-l}^{(2)}$ leads to
a cooperative Jahn-Teller effect. Indeed, neglecting the electron--lattice interaction, the effective Hamiltonian is symmetric in the XY plane, so that Bose condensates with $d_{x^2-y^2}$ and $d_{xy}$ symmetry, or $B_1$ and $B_2$ superconducting modes, are equivalent. This degeneracy is lifted due to electron-lattice interaction with a spontaneous breaking of the XY symmetry with the appearance of local distortions with the most active $B_{1g}$ symmetry and the corresponding condensation of the $B_1$  superconducting mode with $d_{x^2-y^2}\propto B_{1g}$ symmetry.

The many competing order parameters give rise to intrinsic static or dynamic nanoscopic heterogeneity, which, in fact, turns out to be inherent in cuprates\,\cite{Mueller,Pelc}.
Other sources of inhomogeneity are randomly distributed
heterovalent substituents or extra oxygen atoms, which
introduce doped holes or electrons into the copper oxygen planes. In most cases  the
dopants act as charged impurities.

A qualitative, and even more so quantitative, description of both the ground state and the phase diagrams of such systems challenges the theory of condensed matter, which has remained a challenge for more than three decades. The advantage of the spin-pseudospin form of the effective Hamiltonian (\ref{HH}) is that it immediately suggests the use of a simple molecular-field approximation, which has proven itself well for conventional spin-magnetic systems\,\cite{CM_2021}.

Below we consider in more detail several of the most important particular realizations of the phase states of the model Hamiltonian.

\subsection{Atomic limit}
Within "atomic limit" we neglect all the transport terms, so that the Hamiltonian reduces to
\beq
\mathcal{H} =
	 \sum_{i}  (\Delta _{i}{\hat S}_{iz}^2
	  - \mu {\hat S}_{iz}) + \sum_{i>j} V_{ij}{\hat S}_{iz}{\hat S}_{jz}
+ s^2\sum_{i>j}J_{ij} \boldsymbol{\hat \sigma}_i \boldsymbol{\hat \sigma}_j \\
	- \mathbf{h}s \sum_{i} \boldsymbol{\hat \sigma}_i \, ,
\label{Hat}
\eeq
where we have included an additional Zeeman term.
The model generalizes the 2D dilute AFM Ising model with charged
impurities. In the limit $\Delta \rightarrow \infty$ it reduces to the $S$\,=\,1/2 Ising model
with fixed magnetization. At  $\Delta >0$ the results can be compared with
the Blume-Capel model\,\cite{Blume,Capel} or with the Blume-Emery-Griffiths
model\,\cite{BEG}.
This limit of the spin-pseudospin model in the two-sublattice, nearest neighbor  approximation and Ising form for the exchange interaction, was considered in sufficient detail
 in a series of papers\,\cite{Panov_JLTP_2017,Panov_JETPLett_2017,Chikov_2018,Panov_JMMM_19,Panov_JSNM_2019,Panov_PSS_2019}
 within MFA, Bethe cluster approach, and classical Monte-Carlo technique.

The MFA $n$-$\Delta$ phase diagrams for the ground state (GS)  are very different for the case of strong ($\tilde{J}=s^2J>V$) and weak exchange ($\tilde{J}<V$)\,\cite{{Panov_JLTP_2017}}.
In the strong exchange limit the MFA points to the realization of the AFM order given positive $\Delta$ or checkerboard charge order given negative $\Delta$.
However, classical Monte-Carlo calculations for large square lattices show that homogeneous ground state solutions found in the MFA are unstable with respect to phase separation with the charge
and spin subsystems behaving like immiscible quantum liquids\,\cite{Panov_JETPLett_2017}.
With lowering the temperature  the specific heat
 exhibits two successive phase transitions: first, antiferromagnetic ordering in
the spin subsystem diluted by randomly distributed charges, then, the charge condensation in the
charge droplets. It means that the AFM phase in the strong exchange limit is unstable with respect to macroscopic separation of the charge and spin subsystems.
At this point, the AFM matrix pushes out the charges to minimize the surface energy associated with the impurities.
The inhomogeneous droplet phase reduces the energy of the system
and changes the GS phase diagram. Charge doping does
suppress the long-range spin order, but the phase separation (PS) of doped charges and short-range spin order exists for a whole range of the charge doping.

In the weak exchange limit the charged impurities remain distributed randomly over the AFM matrix up to $T=0$, and also the charged impurities remain distributed randomly in the CO phase, as for the near-neighbor interaction the energies of all possible distributions of extra charges over the CO matrix are equal.

The temperature phase diagrams and thermodynamic properties (magnetic susceptibility and specific heat) of the 2D spin-pseudospin system within atomic limit
were studied within MFA, Bethe cluster approach, and classical Monte-Carlo technique
with a special attention given to the role played by the
on-site correlation and doping level\,\cite{Panov_JMMM_19,Panov_JSNM_2019,Panov_PSS_2019}.
It was shown that the competition between the charge
and magnetic orderings leads to the formation of
unusual phase states at finite temperatures.
To describe the thermodynamic properties of inhomogeneous PS state we use the model developed in Ref.\,\cite{Kapcia}.

Critical behavior of the model was studied with  classical Monte Carlo technique on large square lattices with periodic boundary conditions in a "strong" exchange limit\,\cite{Dasha_2020}. In the framework of the finite size scaling theory  the static critical exponents for the specific heat  and the correlation length were obtained for a wide range of the local density-density interaction parameter $\Delta$ and charge density $n$.

\subsection{"Large negative-$U$"\, approximation: NO-CO-BS interplay}


Within "large negative-$U$"\, approximation
 ($\Delta \rightarrow -\infty$)  the energy of the parent [CuO$_4$]$^{6-}$-center becomes infinitely large, so that the on-site Hilbert space is reduced to a doublet $M_S=\pm 1$ of electron and hole centers.
  The system   becomes  equivalent to the well known lattice hard-core ($hc$) Bose system with an inter-site repulsion,
 governed in the nearest-neighbor approximation only by two parameters, $t_b$ and $V$.

 The Hamiltonian of $hc$-bosons on a lattice can be written as follows (see Ref.\,\cite{RMP}  and references therein)
\begin{eqnarray}
	H_{hc}=
		-\sum\limits_{\langle ij \rangle}
	t_{ij} {\hat P} ({\hat b}_{i}^{\dagger}{\hat b}_{j}+{\hat b}_{j}^{\dagger}{\hat b}_{i}) {\hat P} + \sum\limits_{\langle ij \rangle} V_{ij} n_{i} n_{j} - \mu \sum\limits_{i} n_{i},
\label{hcB}
\end{eqnarray}
where ${\hat P}$ is the projection operator which removes double occupancy of
any site, ${\hat b}^{\dagger}({\hat b})$ are
the Pauli creation (annihilation) operators which are Bose-like commuting for
different sites $[{\hat b}_{i},{\hat b}_{j}^{\dagger}]$\,=\,0, if $i\neq j$, but $\{{\hat b}_{i},{\hat b}_{j}^{\dagger}\}$\,=\,0, or $[{\hat b}_{i},{\hat b}_{i}^{\dagger}]=1-2n_i$,
$n_i = {\hat b}_{i}^{\dagger}{\hat b}_{i}$, $N$ is a full number of sites, $\mu $  the chemical potential
determined from the condition of fixed full number of bosons $\sum_{i}\langle n_{i}\rangle $ or concentration $n=\frac{1}{N}\sum_{i}\langle n_{i}\rangle \in [0,1]$. The $t_{ij}$ denotes an effective transfer integral,  $V_{ij}$ is an
intersite interaction between the bosons. For nearest neighbor coupling, $V_{ij}=V_{nn}=V>$\,0, and $t_{ij}=t_{nn}=t>$\,0.
The model of hard-core bosons with an intersite repulsion is
equivalent to a system of $s$\,=\,1/2 spins   exposed to an external magnetic field
in the $z$-direction\,\cite{Matsuda-1970}. For the system with neutralizing background we arrive at an effective pseudospin $s$\,=\,1/2  Hamiltonian
\begin{equation}
H_{hc} = \sum_{\langle ij \rangle} J^{xy}_{ij} ({\hat s}_{i}^{+}{\hat s}_{j}^{-}+{\hat s}_{j}^{+}{\hat s}_{i}^{-}) +\sum\limits_{\langle ij \rangle}
J^{z}_{ij} {\hat s}_{i}^{z} {\hat s}_{j}^{z} - \mu \sum\limits_{i} {\hat s}_{i}^{z},
\label{spinB}
\end{equation}
where
$J^{xy}_{ij}=2t_{ij}$, $J^{z}_{ij}=V_{ij}$,
$\hat{s}_i^{-}= \frac{1}{\sqrt{2}}\hat{b}_i $, $\hat{s}_i^{+}=-\frac{1}{\sqrt{2}}\hat{b}_i^{\dagger}$, $\hat{s}_i^{z}=-\frac{1}{2}+\hat{b}_i^{\dagger}\hat{b}_i$,
$\hat{s}_i^{\pm}=\mp \frac{1}{\sqrt{2}}(\hat{s}_i^x \pm i\hat{s}_i^y)$.

Local on-site order is characterized by the three order parameters: $\langle {\hat s}^z\rangle$\,=$n$\,-\,$\frac{1}{2}$; $\langle {\hat s}^{\pm}\rangle$\,=\,$|\langle {\hat s}^{\pm}\rangle |e^{\pm i\varphi}$, related with the charge and superfluid degree of freedom, respectively.

In general, depending on the relative values of the parameters the phase diagram of the system is shown to consist of at least seven different states, including four uniform phases (NO, CO,  BS, and a supersolid, or mixed CO-BS phase) and 3 types of phase separated (PS) states: CO-BS (PSI), CO-NO-normal (PS2) and the state of charge droplets (PS3)\,\cite{Rob}.

In addition to the conventional charge-ordering (CO)  and Bose superfluid (BS) phases the MFA predicts the appearance of an unconventional uniform supersolid phase (SS) with the "on-site"\, coexistence of the  insulating and  superconducting properties\,\cite{RMP}.
 However, detailed analysis of the hard-core boson model shows that the supersolid phase is a mean-field artefact, de facto this homogeneous phase is intrinsically unstable. For $nn$-interactions and finite temperatures the phase-separated CO-BS phase  has a lower energy than the uniform SS phase\,\cite{Rob,Panov_PSS_2019}, though at $T$\,=\,0 the both phases have the same energy.
The CO-BS phase separation is also confirmed by more accurate calculations by the quantum Monte-Carlo method\,\cite{Schmid2002}.

The analogy with the $s$\,=\,1/2 spin system allows one to use many of the qualitative results known for quantum magnets\,\cite{Manousakis}.
So,  the true ground state of the  $s$\,=\,1/2 antiferromagnet (given even number of spins) is a quantum superposition of all possible states with full spin $S$\,=\,0 and
zero value of the local order parameter: $\langle {\bf s}_i\rangle$\,=\,0. The N\'eel state is just a classic "component"\, of this "hidden"\, quantum state,
so-called "physical"\, ground state. The contribution of purely quantum states is manifested in a significant decrease (up to 40\%) in the value of the local order parameter
in the N\'eel "portrait"\, as compared with the nominally maximum value of $s$\,=\,1/2.
A similar situation will be observed in the $hc$-boson model.

Making use of a special algorithm for CUDA architecture for NVIDIA graphics cards that implies a nonlinear conjugate-gradient method to minimize energy functional and Monte-Carlo technique we have been able to directly observe formation of the ground state configuration for the 2D hard-core bosons  with lowering the temperature and its transformation with increase the temperature and boson concentration\,\cite{Moskvin-JSNM-2017,Moskvin-JSNM-2018,Moskvin-JSNM-2019}, allowing us to examine earlier  implications\,\cite{bubble,bubble-2} and uncover novel features of the phase transitions, in particular, look upon the nucleation of the odd domain structure, the localization of the bosons doped away from half-filling, and the phase separation regime.
Interestingly, the accuracy of numerical calculations has been limited making it possible to reproduce the effect of  minor inhomogeneities common to any real crystal.

Typically for small and moderate anisotropy the annealing is finished by formation a system of domains with closed-loop domain walls which quickly collapse thus setting an uniform single-domain CO ground state with a hardly noticeable remnant inhomogeneity. At the lowest temperatures we can form an almost ideal charge ordered checkerboard structure at half-filling that does not modify with increasing the temperature up to $T_{CO}$.

However, systematic studies have indicated that in  some cases there occurs a low-temperature CO domain structure with stable stripe-like disconnected (within our lattice size)  domain walls oriented along main lattice axes. Along with a simple uniform ("ferromagnetic") SF phase parameter distribution these 1D walls can have unconventional multidomain topological structure of the SF phase order parameter with a high density of 2$\pi$ domain walls separating the 1D phase domains.

Thus, the features of phase separation in real materials will depend on many, sometimes even the smallest, inhomogeneities of the initial states.

However, the "large negative-$U$"\, model with $\Delta \rightarrow \infty$, despite all the allusions, is hardly directly related to HTSC cuprates.
Effect of finite on-site correlations for a simple effective model of a superconductor with very short coherence length in which electrons are localized and only electron pairs have a possibility of transferring was considered by Kapcia {\it et al.}\cite{Kapcia}. The model is a
simple generalization of the standard model of non-interacting local composite bosons to the case of finite pair binding energy $U$.
 The phase diagrams and thermodynamic properties of the model have been determined within the
effective field theory for different doping and  chemical potential.  Depending on the values of interaction parameters, the system can exhibit not only the homogeneous phases, superconducting (BS) and nonordered (NO), but also the phase separated states (PS: BS-NO). The system considered exhibits interesting multicritical behaviour including tricritical points.

The discovery of superconductivity in undoped $T^{\prime}$-cuprates R$_2$CuO$_4$ (R = Pr, Nd, Sm, Eu, and Gd) with no apical oxygen\,\cite{Naito-2016} challenges the long-thought understanding that the
parent compounds of the high-$T_c$ cuprates for sure are AF Mott insulators and casts doubt on the leading HTSC scenario,  which links the appearance of superconductivity with the pairing of doped carriers indicating a situation close to that realized in the local boson model.

It is obvious that a small positive, especially negative, value of the local correlation energy $U_{th}$, or the 11-20 (02) charge transfer energy, in undoped $T^{\prime}$-cuprates leads to the production of metastable EH dimers, which are the progenitors of several new phases.
The quantum EH dimers in CuO$_2$ planes of a parent
cuprate can give rise to a homogeneous quantum phase of
the type of a quantum spin liquid that is the superposition of the states of the type of resonating valence bonds (RVB)\,\cite{RVB}. It is
evident that, at some critically low energy $U_{th}$, it is the
quantum liquid of EH dimers, rather than an antiferromagnetic insulator, that can be the ground state of the parent $T^{\prime}$-cuprate. Strictly speaking, the  quantum liquid is characterized by zero local order parameters and does not resemble typical N\'eel-like phases with nonzero local order parameters such as checkerboard charge order, or some magnetic order.

In general, at rather large negative values of the on-site correlation parameter $U_{th}$ we should consider the CuO$_2$ planes in the  $T^{\prime}$-cuprates to be a strongly correlated system of electron [CuO$_4^*$]$^{7-}$ and hole [CuO$_4^*$]$^{5-}$ centers coupled by inter-site correlations and two-particle transport.
However, the EH dimers can be progenitors of N\'eel-like phases such as a charge order  and superconducting phase with a long-range order of local order parameters.
The N\'eel phases start to form at high temperatures in the nonordered phase, when thermal fluctuations and fluctuating non-uniform fields  destroy the energetically preferred quantum states, while the N\'eel-type domains become more and more extended and stable with decreasing the temperature,  leaving no real chance of the formation of a true quantum ground state in the low-temperature limit.
 The final N\'eel-type state even though the local order parameters may be reduced,  is stable against quantum fluctuations at T\,=\,0.
Similar effects are expected for N\'eel phases (AFMI, CO, BS) in T$^{\prime}$-cuprates.

Parent $T$-cuprates are antiferromagnetic insulators with a low concentration of metastable EH
dimers at finite temperatures,  however, under the nonisovalent substitution accompanied by strong suppression of the thermal CT gap and EH bonding energy
  these transform into bosonic superconductors.

\subsection{Electron-hole interplay in "normal"\, metallic state}
We believe that both the antiferromagnetic (AFMI) phase, charge order (CO), bosonic superconductivity (BS), and the Fermi metallic (FL) phase can be phase states both of the doped and parent cuprate.
In other words, we can consider the coherent metallic state of our model cuprate by omitting the spin exchange interaction and the transfer of composite bosons.
 In this case, the ground state of the cuprate is associated with the half-filled 2D band. In the crudest single-band approximation, the Fermi surface for
this band is an array of squares touching at the corners, which can be regarded as containing
electrons around $\Gamma$-point (0,0) or containing holes around $X$-point ($\pi$,$\pi$). Such a
Fermi surface has perfect nesting for the wave vector
${\bf Q}$\,=\,$X$\,=\,($\pi$, $\pi$). The nesting can lead to either a
charge-density wave, which for this commensurate wave
vector could show up as an $X$-point soft-mode transition,
or a spin-density wave, or possibly both. In fact, candidates for both of these instabilities occur, leading to a
considerable amount of analysis and speculation about
Fermi-surface-driven instabilities in the La214 system.


Strictly speaking, elementary excitations over the ground state, that is electrons and holes, should be described by Hamiltonian ${\hat H}_{FL}$
\beq
{\hat H}_{FL}=\Delta -\mu \sum_{i\nu}(p_{i\nu}-n_{i\nu}) +{\hat V}_{int} +{\hat H}_{kin}^{(1)} \, ,
\eeq
\beq
{\hat V}_{int}=\sum_{i>j}\sum_{\nu}V_{ij}(p_{i\nu}p_{j\nu}+n_{i\nu}n_{j\nu}-2p_{i\nu}n_{j\nu}) \, ,
\eeq
where $\Delta $ is a gap due to local correlations, $\mu$ is the chemical potential,
which depends on the average number of holes/electrons, $p_{i\nu}={\hat P}_{i+}^{\nu}{\hat P}_{i-}^{\nu}$ and $n_{i\nu}={\hat N}_{i-}^{\nu}{\hat N}_{i+}^{\nu}$ are the on-site occupation numbers for holes and electrons, respectively.
For simplicity, hereafter we'll neglect nonlocal correlations.
After Fourier transformation we arrive at a new form of the Hamiltonian  ${\hat H}_{kin}^{(1)}$ (\ref{Hkin1}) as follows
$$
{\hat H}_{kin}^{(1)}= \sum_{{\bf k}\nu}[\epsilon^p_{{\bf k}}{\hat P}_{{\bf k}+}^{\nu}{\hat P}_{{\bf k}-}^{\nu}+
 \epsilon^n_{{\bf k}}{\hat N}_{{\bf k}+}^{\nu}{\hat N}_{{\bf k}-}^{\nu}+
 \frac{1}{2} \epsilon^{pn}_{{\bf k}}({\hat P}_{{\bf k}+}^{\nu}{\hat N}_{{\bf k}-}^{\nu}+{\hat P}_{{\bf k}-}^{\nu}{\hat N}_{{\bf k}+}^{\nu}) +h.c.] =
$$
 \begin{equation}
 \sum_{{\bf k}\nu}{\hat \Psi}_{{\bf k}\nu}^{\dag}{\hat H}_{{\bf k}}{\hat \Psi}_{{\bf k}\nu} \, ,
\label{H1k}	
\end{equation}
where  operators ${\hat P}_{{\bf k}-}^{\nu}$ and 	${\hat N}_{{\bf k}-}^{\nu}$ are  gathered in the two-component operator ${\hat \Psi}_{{\bf k}\nu}$, so that the Hamiltonian takes the form of a 2$\times$2 matrix
\begin{equation}
{\hat \Psi}_{{\bf k}\nu}=
\begin{pmatrix}
	{\hat P}_{{\bf k}-}^{\nu} \\
	{\hat N}_{{\bf k}-}^{\nu}
	\end{pmatrix},
{\hat H}_{{\bf k}}=\begin{pmatrix}
	\epsilon^{p}_{{\bf k}}&\epsilon^{pn}_{{\bf k}} \\
	\epsilon^{np}_{{\bf k}}&\epsilon^{n}_{{\bf k}}
	\end{pmatrix}
\end{equation}
with
\begin{equation}
\epsilon^p_{{\bf k}}=\sum_{i\not=j}e^{i{\bf k}{\bf R}_{ij}}t^p_{ij} \,,\epsilon^n_{{\bf k}}=\sum_{i\not=j}e^{i{\bf k}{\bf R}_{ij}}t^n_{ij} \,,
\epsilon^{pn}_{{\bf k}}=\sum_{i\not=j}e^{i{\bf k}{\bf R}_{ij}}t^{pn}_{ij}\,,\epsilon^{np}=(\epsilon^{pn})^* \,.
\end{equation}
All  these "band" parameters have a similar structure, so that taking into account the interaction of the first, second, third and fourth neighbors in CuO$_2$ planes, these can be represented as follows
$$
\epsilon_{{\bf k}}^{p,n,pn}=-2t_1^{p,n,pn}(\cos k_x+\cos k_y)+
$$
$$
4t_2^{p,n,pn}\cos k_x\cos k_y-2t_3^{p,n,pn}(\cos2k_x+\cos2k_y)-
$$
\beq
4t_4^{p,n,pn}(\cos2k_x\cos k_y+\cos2k_y\cos k_x)  \, ,
\eeq
where  wave vectors $k_x$ and $k_y$ are in units $\pi$/a. As noted above, the parameters of the correlated transport $t_i^{p,n,pn}$\,(i=1, 2, 3, 4, ...) can be very different, although this circumstance is ignored in the overwhelming majority of theoretical models based on representations of the simple single-electron tight-binding model.
Generally accepted values of the tight-binding band parameters are as follows\,\cite{Andersen}:
$t_1\approx$\,0.4\,eV, $t_2\approx$\,0.3\,$t_1$, and $t_3\approx$\,0.5\,$t_2$. However, more later data\,\cite{Mark} and most recent calculations point to a more significant contributions of the third and fourth neighbors with $t_2/t_1$\,=\,0.136,  $t_3/t_1$\,=\,0.068, $t_4/t_1$\,=\,0.061, where $t_1$\,=\,0.28\,$t_{pd}$ ($t_{pd}\simeq$\,1.2--1.5\,eV)\,\cite{Photo}. Despite all the limitations of the simple model, tight-binding parameters can be used to estimate the true parameters of correlated transport.

If $t_{pn}$\,=\,0 then at relevant dopings $x$ the Fermi surface consists of one large hole surface centered at the $X$ point ($\pi ,\pi$). If
$t_{pn}\not=$\,0 then the Fermi surface is reconstructed. The details
depend on the band filling and the magnitude of $t_{pn}$.

Making use of a unitary transformation
\begin{equation}
{\hat E}_{{\bf k}}={\hat U}_{{\bf k}}{\hat H}_{{\bf k}}{\hat U}_{{\bf k}}^{\dag}\,,{\hat U}_{{\bf k}}=\begin{pmatrix}
	u_{{\bf k}}&v_{{\bf k}} \\
	-v^*_{{\bf k}}&u^*_{{\bf k}}
	\end{pmatrix}
\end{equation}
we arrive at a diagonalized form of the Hamiltonian
\beq
{\hat H}_{kin}^{(1)}=\sum_{{\bf k}\nu}{\hat \Phi}_{{\bf k}\nu}^{\dag}{\hat E}_{{\bf k}}{\hat \Phi}_{{\bf k}\nu}
\eeq
with
\beq
{\hat \Phi}_{{\bf k}\nu}={\hat U}_{{\bf k}}{\hat \Psi}_{{\bf k}\nu}=\begin{pmatrix}
	u_{{\bf k}}{\hat P}_{{\bf k}-}^{\nu}+v_{{\bf k}}{\hat N}_{{\bf k}-}^{\nu}  \\
	u^*_{{\bf k}}{\hat N}_{{\bf k}-}^{\nu}-v^*_{{\bf k}}{\hat P}_{{\bf k}-}^{\nu}
	\end{pmatrix}, {\hat E}_{{\bf k}}=\begin{pmatrix}
	\epsilon_{{\bf k}+}&0 \\
	0&\epsilon_{{\bf k}-}
	\end{pmatrix},
\eeq
where
\beq
\epsilon_{{\bf k}\pm}=\frac{1}{2}\left(\epsilon^p_{{\bf k}}+\epsilon^n_{{\bf k}}\pm \sqrt{(\epsilon^p_{{\bf k}}-\epsilon^n_{{\bf k}})^2+4|\epsilon^{pn}_{{\bf k}}}|^2\right) \,.
\eeq
Obviously, disregarding the $PN$ term, we are dealing with the bands of noninteracting $P$- and $N$-modes, that is, holes and electrons. The inclusion of the electron-hole interaction of the $PN$-type  ("creation/annihilation of an electron -- annihilation/creation of a hole") leads to the quantum renormalization of modes due to $P$-$N$ superpositions with the appearance of a ${\bf                           k}$-dependent gap and a significant rearrangement of the band structure with the Fermi surface reconstruction and specific electron-hole interplay, in particular, in the Hall effect.
It should be emphasized that the renormalized bands and  the semimetal-like band structure can be very sensitive to the
choice of the correlated charge transfer parameters.

On the other hand
Fermi-like operators ${\hat P}_{+}^{\nu}$/${\hat P}_{-}^{\nu}$ and ${\hat N}_{-}^{\nu}$/${\hat N}_{+}^{\nu}$ can be interpreted to be creation/annihilation operators for holes and electrons ${\hat P}_{\nu}^{\dag}$/${\hat P}_{\nu}$ and ${\hat N}_{\nu}^{\dag}$/${\hat N}_{\nu}$, respectively, acting on  ground state formed by the half-filled band of parent cuprate.

In terms of novel operators the  Hamiltonian ${\hat H}_{kin}^{(1)}$ reads as follows:
\begin{equation}
{\hat H}_{kin}^{(1)}= -\sum_{i>j}\sum_{\nu} [t^p_{ij}{\hat P}_{i\nu}^{\dag}{\hat P}_{j\nu}+
 t^n_{ij}{\hat N}_{i\nu}^{\dag}{\hat N}_{j\nu}+
 \frac{1}{2} t^{pn}_{ij}({\hat P}_{i\nu}^{\dag}{\hat N}_{j\nu}^{\dag}+{\hat P}_{i\nu}{\hat N}_{j\nu}) +h.c.] \,,
\label{H1}	
\end{equation}
or after Fourier transformation
\begin{equation}
{\hat H}_{kin}^{(1)}= \sum_{{\bf k}\nu}[\epsilon^p_{{\bf k}}{\hat P}_{{\bf k}\nu}^{\dag}{\hat P}_{{\bf k}\nu}+
 \epsilon^n_{{\bf k}}{\hat N}_{{\bf k}\nu}^{\dag}{\hat N}_{{\bf k}\nu}+
 \frac{1}{2} \epsilon^{pn}_{{\bf k}}({\hat P}_{{\bf k}\nu}^{\dag}{\hat N}_{-{\bf k}\nu}^{\dag}+{\hat P}_{-{\bf k}\nu}{\hat N}_{{\bf k}\nu}) +h.c.
 ] .
\label{H11}	
\end{equation}

The alternative form of the Hamiltonian ${\hat H}_{kin}^{(1)}$ (\ref{H11}) directly indicates the connection between the $PN$ terms and the operators of creation and annihilation of electron-hole pairs with the opposite direction of the quasimoments of the electron and hole.

Now there is strong evidence  that the electron- and
hole-like charge carriers coexist in a broad range of
superconducting cuprates, including both the nominally $n$- and $p$-type. The coexistence of electrons and holes is necessary to explain the
normal state resistivity and Hall effect (see, e.g., Refs.\,\cite{Harshman,Luo}).

Finally, to effect strong electron-hole coupling, the group velocity
of the relevant carriers need to be close to one
another, or near zero. The flat saddle point at $(0,\pi )$, $(\pi ,0)$
is an ideal harbor for such a coupling\,\cite{Luo}.



\section{Conclusion}
In summary, we have presented an unified non-BCS approach to the description of the high-$T_c$ cuprates based on a minimal model for the CuO$_4$-centers in CuO$_2$ planes  with the on-site Hilbert space  reduced to only three effective valence centers [CuO$_4^*$]$^{7-}$, [CuO$_4$]$^{6-}$, [CuO$_4^*$]$^{5-}$, or cluster analogs of Cu$^{1+,2+,3+}$ centers, forming a "well isolated"\, charge triplet.
Instead of conventional quasiparticle $\bf k$-momentum description we made use of a real space on-site "unparticle"\, $S$\,=\,1 pseudospin formalism to describe the charge triplets and introduce an  effective spin-pseudospin Hamiltonian which takes into account main on-site and inter-site interactions, as well as electron-lattice effects. We argue that antiferromagnetic insulating, charge ordered, superconducting, and Fermi-liquid phases are possible phase states of a model parent cuprate. Typical phase state of a doped cuprate, in particular mysterious pseudogap phase, is a result of a phase separation\,\cite{CM_2021}.
Superconductivity of cuprates is not a consequence of pairing of doped holes, but the result of  quantum transport of on-site composite hole bosons, whereas main peculiarities of normal state can be related both to an electron-hole interplay for unusual Fermi-liquid phase and features of the phase separation\,\cite{CM_2021}.
We believe that the $d_{x^2-y^2}$ symmetry of the local superconducting order parameter is related with strong on-site electron-lattice coupling with the rhombic $Q_{B_{1g}}$-mode.

This research was supported by the Ministry of Education and Science, project No FEUZ-2020-0054.

\end{document}